\documentclass[prl,aps,superscriptaddress,twocolumn,showpacs]{revtex4}
\usepackage{amsfonts}
\usepackage{amsmath}
\usepackage{amssymb}
\usepackage{graphicx}
\usepackage{color}

\begin{document}

\title{Dissipative structures in optomechanical cavities}
\author{Joaqu\'{\i}n Ruiz-Rivas}
\affiliation{Departament d'\`Optica, Universitat de Val\`encia, Dr. Moliner 50, 46100
Burjassot, Spain}
\author{Carlos Navarrete-Benlloch}
\affiliation{Max-Planck-Institut f\"ur Quantenoptik, Hans-Kopfermann-Str. 1, D-85748
Garching, Germany}
\author{Giuseppe Patera}
\affiliation{Laboratoire PhLAM, Universit\'{e} Lille 1, 59655 Villeneuve d'Ascq Cedex,
France}
\author{Eugenio Rold\'{a}n}
\affiliation{Departament d'\`Optica, Universitat de Val\`encia, Dr. Moliner 50, 46100
Burjassot, Spain}
\author{Germ\'{a}n J. de Valc\'{a}rcel}
\affiliation{Departament d'\`Optica, Universitat de Val\`encia, Dr. Moliner 50, 46100
Burjassot, Spain}

\begin{abstract}
Motivated by the increasing interest in the properties of multimode
optomechanical devices, here we study a system in which a driven mode of a
large-area optical cavity is despersively coupled to a deformable mechanical
element. Two different models naturally appear in such scenario, for which
we predict the formation of periodic patterns, localized structures (cavity
solitons), and domain walls, among other complex nonlinear phenomena.
Further, we propose a realistic design based on intracavity membranes where
our models can be studied experimentally. Apart from its relevance to the
field of nonlinear optics, the results put forward here are a necessary
step towards understanding the quantum properties of optomechanical systems
in the multimode regime of both the optical and mechanical degrees of
freedom.
\end{abstract}

\maketitle

\textbf{Introduction.} Since the eighties of the past century, the advent
and rapid growing of quantum information have boosted the research on
quantum technologies that try to develop microdevices allowing for robust
implementations of controllable quantum interactions for long coherence
times. By today the state of the art permits the fabrication of devices
functioning in such strong coupling regime with unprecedented control and
accuracy in a rapidly growing field. These new quantum devices are very
diverse including cavity--QED \cite{CQED}, optical lattices \cite{OL},
trapped ions \cite{TI}, superconducting circuits \cite{SC}, quantum dots 
\cite{QD}, atomic ensembles \cite{AE}, etc.

In typical optomechanical cavities \cite{OM} the interaction occurs between
a light field and a mechanical oscillator via radiation pressure. This type
of system has been known for a long time in classical optics \cite{Meystre},
and to some extent resembles a nonlinear Kerr cavity \cite{LL} as the cavity
length (and consequently the cavity resonance) depends on the intracavity
field intensity. At present, the stress is made on their capability to show
quantum coherent phenomena such as cooling and amplification \cite{CoolAmp},
strong (linear) coupling effects like optomechanically induced transparency 
\cite{OMIT}, or to prepare squeezed states of light dissipatively \cite%
{Fabre94,Mancini94,Brooks12,Safavi13}. Attention has also been recently paid
to the nonlinear dynamics of optomechanical arrays \cite{OMA}, cavities in
which radiation pressure competes with the photothermal effect \cite{RPPT},
planar dual-nanoweb waveguides subject to radiation pressure \cite{DualNano}%
, and optomechanical cavities containing atomic ensembles \cite{Atomic}.

Except for some of these recent works, up to now most studies deal with a
small number of modes either in the optical or the mechanical degrees of
freedom, while the nonlinear interplay between many optical and mechanical
modes entails the existence of correlations among them that can provide
optomechanical systems with new capabilities. At the quantum level, these
correlations may lead to multipartite entanglement \cite%
{Horodecki,Pfister,Patera12}; at the classical level, they help for the
spontaneous appearance of dissipative structures that are long range ordered
configurations, including periodic, quasiperiodic and aperiodic patterns, as
well as localized structures, which may also exhibit nontrivial temporal
behavior \cite{StaliunasVSM}.

The interplay between the quantum and classical perspectives in multimode
systems has received theoretical attention mainly in the context of optical
parametric oscillators for which some exciting phenomena were predicted \cite%
{OPOs}. No such studies exist concerning optomechanical systems, and here we
make a step towards this goal by proposing a multimode optomechanical cavity
configuration in which dissipative structures are predicted to appear. We
keep our study at the classical level by concentrating on the analysis of
the general conditions under which such patterns can appear and how they
are, leaving the study of their quantum properties for a future publication.
As we show below, periodic patterns, localized structures (cavity solitons), and
domain walls are predicted to occur in a wide region of the two models that
naturally follow from the proposed configuration.

\begin{figure}[t]
\includegraphics[width=0.75\columnwidth]{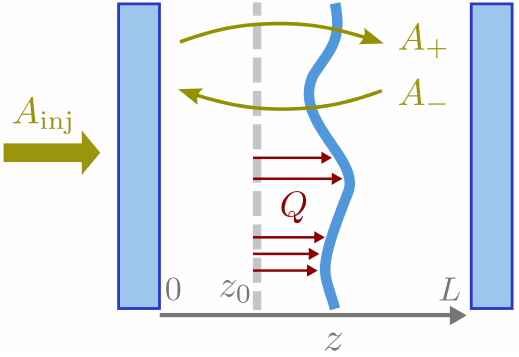}
\caption{Cartoon of the system.}
\label{FigScheme}
\end{figure}

\textbf{Model.} Consider an optical cavity with large-area mirrors
containing a dispersive element which can be thought of as a thin tense
membrane that can be deformed locally, but is also allowed to oscillate as a
whole (Fig. \ref{FigScheme}). Energy is fed in the cavity from the outside
by injecting a coherent laser field through the partially transmitting
mirror, modelled as a paraxial beam%
\begin{equation}
E_{\mathrm{inj}}\left( z,\mathbf{r},t\right) =\mathrm{i}\mathcal{V}A_{%
\mathrm{inj}}\left( z,\mathbf{r},t\right) e^{\mathrm{i}\left( k_{\mathrm{L}%
}z-\omega _{\mathrm{L}}t\right) }+\mathrm{c.c.,}
\end{equation}%
where $\mathbf{r}=\left( x,y\right) $ denotes the position in the plane
transverse to the cavity axis ($z$-axis), and $\mathcal{V}$ is a constant
having the dimensions of voltage, which we choose as $\mathcal{V}=\sqrt{%
\hbar \omega _{\mathrm{c}}/4\varepsilon _{0}L}$ in order to make contact
with quantum optics (see \cite{SupMat,OPOs}), $\omega _{\mathrm{c}}$ being
the frequency of the longitudinal cavity mode closest to the injected
frequency $\omega _{\mathrm{L}}$ (with corresponding wave vectors $%
k_{j}=\omega _{j}/c$ and wavelengths $\lambda _{j}=2\pi /k_{j}$). The
intracavity field $E\left( z,\mathbf{r},t\right) $ can be written
generically as%
\begin{equation}
E\left( z,\mathbf{r},t\right) =\mathrm{i}\mathcal{V}\left( A_{+}e^{\mathrm{i}%
k_{\mathrm{L}}z}+A_{-}e^{-\mathrm{i}k_{\mathrm{L}}z}\right) e^{-i\omega _{%
\mathrm{L}}t}+\mathrm{c.c.,}
\end{equation}%
which is the superposition of two waves with slowly varying complex
amplitudes $A_{\pm }\left( z,\mathbf{r},t\right) $, propagating along the
positive ($A_{+}$) and negative ($A_{-}$) $z$ direction. We first derive an
evolution equation for the amplitude $A_{+}\left( z=L,\mathbf{r},t\right) $
at the end mirror's surface, which we will denote by $A\left( \mathbf{r}%
,t\right) $. The procedure consists in propagating the field along a full
cavity roundtrip \cite{SupMat,hyperbolic}, which, in the paraxial approximation,
and assuming that the fields change very little after one roundtrip (with
associated time $t_{\mathrm{c}}=2L/c$), as well as a small reflectance $\varrho
^{2}$ of the membrane, leads to \cite{SupMat,hyperbolic,BS,SI}%
\begin{eqnarray}
\partial _{t}A &=&\gamma _{\mathrm{c}}\mathcal{E}+\gamma _{\mathrm{c}}\left( 
\mathcal{-}1+\mathrm{i}\Delta +\mathrm{i}l_{\mathrm{c}}^{2}\nabla _{\bot
}^{2}\right) A  \label{dAdtaux} \\
&&-\mathrm{i}\frac{4\gamma _{\mathrm{c}}\varrho }{T}\cos \left[ 2k_{\mathrm{L%
}}\left( z_{0}+Q\right) \right] A  \notag
\end{eqnarray}%
where $T=T_{1}+T_{2}$ is the sum of the mirrors' transmittances, $\gamma _{%
\mathrm{c}}=cT/4L\ $is the cavity damping rate, $\Delta =\left( \omega _{%
\mathrm{L}}-\omega _{\mathrm{c}}\right) /\gamma _{\mathrm{c}}$ is the
dimensionless detuning parameter, $l_{\mathrm{c}}^{2}=2L/k_{\mathrm{L}}T$ is
the diffraction length (squared), $\nabla _{\bot }^{2}=\partial
_{x}^{2}+\partial _{y}^{2}$ is the transverse Laplacian, and $\mathcal{E} =2\sqrt{T_{1}/T^{2}}A_{\mathrm{inj}}$ is a scaled version of the injection
field amplitude. The local displacement of the membrane perpendicular to its
rest position at $z=z_{0}$ is measured by the field $Q\left( \mathbf{r}%
,t\right) $, so $Q=0$ in the absence of illumination.

The equation for the light field must be complemented with an equation for
the displacement field $Q(\mathbf{r},t)$. In the absence of forcing and
dissipation $Q$ obeys a wave equation $\partial _{t}^{2}Q+\mathbb{L}Q=0$,
where $\mathbb{L}$ is a suitable differential operator whose form depends on the specific implementation. Including damping, assumed homogeneous for
simplicity, and coupling of the membrane to the light field, we get \cite{SupMat}
\begin{equation}
\partial _{t}^{2}Q+\gamma _{\mathrm{m}}\partial _{t}Q+\mathbb{L}Q=\frac{%
4\varrho \hbar k_{\mathrm{L}}}{\sigma t_{\mathrm{c}}}\sin \left[ 2k_{\mathrm{%
L}}\left( z_{0}+Q\right) \right] \left\vert A\right\vert ^{2},  \label{ecQ}
\end{equation}
where $\sigma $ is the membrane's surface mass density.

For the sake of simplicity and analyticity, we choose $\mathbb{L}=\Omega _{\mathrm{m}}^{2}-v^{2}\nabla _{\bot }^{2}$, which models a membrane (characterized by
its sound speed $v$) that can oscillate as a whole at frequency $\Omega _{\mathrm{m}}$. This choice ensures that the model will possess spatially homogeneous solutions (invariant under translations across the transverse plane), which can become unstable in favor of patterns that break such symmetry, as usual in spontaneous pattern forming scenarios \cite{CrossHohenberg}. As we show below through numerical simulations, the existence of such sufficiently homogeneous solutions turns out to be essential for pattern formation, and later we propose a specific experimental implementation leading to this particular choice of $\mathbb{L}$.

\begin{figure*}[t]
\centering
\includegraphics[width=0.99\textwidth]{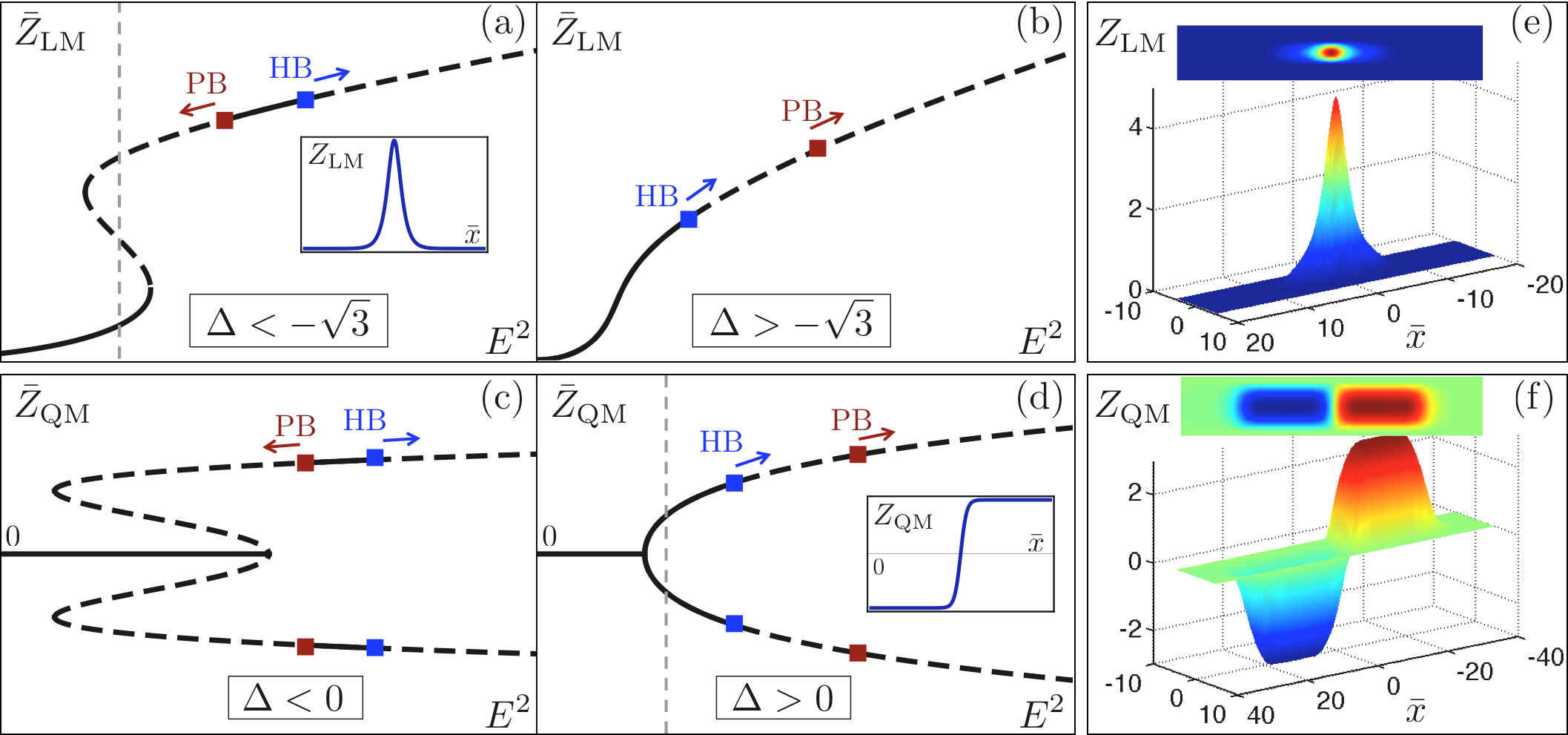}
\caption{(a-d) Cartoon of the generic steady-states and instabilities of the
linear (LM) and quadratic (QM) models introduced in the text. We plot the mechanical
displacement $\bar{Z}$ of the homogeneous steady-state solution as a
function of the injection $E^{2}$, with the stable (unstable) regions
denoted by solid (dashed) lines. We mark the Pitchfork bifurcation leading
to pattern formation (red squares), as well as the Hopf instability (blue
squares) leading to oscillatory solutions, the arrows indicating the portion
of the curve which they make unstable. The insets shown in (a) and (d) show the type of patterns expected for injections marked by the thin grey line, corresponding to solitons in the linear model and domain walls in the quadratic one. (e,f) Localized structures found by numerically
solving the equations modelling a realistic implementation based on framed membranes, expected to lead to an effective 1D model with homogeneous mode and patterns occurring along the $x$ axis, as long as only one or few modes of the membrane are excited in the $y$ direction; in (e) we show a solitonvappearing in the region marked with a vertical dashed line in (a), while in (f) we show a domain wall appearing in the region marked in (d). The specific parameters of the simulation can be checked in \cite{SupMat}.}
\label{Fig2}
\end{figure*}

Eqs. (\ref{dAdtaux}) and (\ref{ecQ}) constitute the basic model for our
optomechanical setup, which coincides with previous single-mode models \cite{Thompson08,Jayich08} in
the limit of small reflectance $\varrho ^{2}$ and ignoring transverse spatial effects.
The type of model that appears depends strongly on the location $z_{0}$ of
the membrane. In this work we consider two typical situations leading to
significantly different scenarios \cite{Thompson08,Jayich08}: either the membrane is located at a node
of the driven mode's standing wave ($z_{0}/\lambda _{\mathrm{L}}=0,1/2,1,...$%
) or half-way between a node and an antinode ($z_{0}/\lambda _{\mathrm{L}%
}=1/8,3/8,5/8,...$). We will refer to the corresponding models as the 
\textit{quadratic} and \textit{linear} models, respectively, because this is the
dependence that the light's frequency shift has on the mechanical field on
each case, see Eq. (\ref{dF}) below. We simplify the model equations in these two
configurations by following the usual procedure in which, taking into
account that the membrane's displacement $Q$ is small compared with the
laser's wavelength, the coupling terms in Eqs. (\ref{dAdtaux}) and (\ref{ecQ}%
) can be expanded around $Q=0$ to the first nontrivial order in $Q$. The two
resulting models can be written in a compact and clean form by introducing
an index $\mu$ which equals 1 or 2 for the linear or quadratic models, respectively, as well as normalized variables and parameters. Defining the
dimensionless time $\tau =\gamma _{\mathrm{c}}t$, spatial coordinates $%
\mathbf{\bar{r}}=\mathbf{r}/l_{\mathrm{c}}$, plus parameters $\gamma =\gamma
_{\mathrm{m}}/\gamma _{\mathrm{c}}$ and $\Omega =\Omega _{\mathrm{m}}/\gamma
_{\mathrm{c}}$, we get \cite{SupMat} 
\begin{subequations}
\label{model_norm}
\begin{gather}
\partial _{\tau }F=\left[ -1+\mathrm{i}\left( \Delta _{\mu }+\nabla
^{2}+Z^{\mu }\right) \right] F+E,  \label{dF} \\
\partial _{\tau }^{2}Z+\gamma \partial _{\tau }Z+\Omega ^{2}\left( 1-\rho
^{2}\nabla ^{2}\right) Z=\Omega ^{2}Z^{\mu -1}\left\vert F\right\vert ^{2},
\label{dZ}
\end{gather}
\end{subequations}
where $F$, $Z$, and $E$ are normalized dimensionless versions of the optical
field $A$, the mechanical field $Q$, and the pump $\mathcal{E}$,
respectively \cite{SupMat}, $\nabla ^{2}=\left( \partial _{\bar{x}%
}^{2}+\partial _{\bar{y}}^{2}\right) $, $\Delta _{\mu }=\Delta - 4(\mu
-1)\varrho /T$, and we have defined the `effective rigidity' parameter $\rho =v/\Omega _{\mathrm{m}}l_{\mathrm{c}}$,
so that the larger $\rho $ the more rigidly the membrane behaves.
Alternatively, $\rho $ is a measure of how local the response of the
membrane is to a stress: when $\rho \rightarrow 0$ the response is local,
while for $\rho \rightarrow \infty $ it is completely integrated, that is,
only the homogeneous mode is excited.

Let us remark that we have also considered the situation in which the end
mirror is the deformable mechanical element, optomechanical coupling arising
from the radiation pressure force \cite{OM} in such case. The resulting
normalized model coincides with the linear model introduced above, Eq. (\ref%
{model_norm}) with $\mu =1$. Also, note that we have not included the case
of the membrane being in an antinode ($z_{0}/\lambda _{\mathrm{L}%
}=1/4,3/4,5/4,...$), what results in a model like the one above with $\mu=2$ but with an extra negative sign in front of the terms $Z^\mu$ and $Z^{\mu-1}$, because we haven't found any pattern forming instabilities in such configuration \cite{SupMat}.

\textbf{Homogeneous solutions, their stability, and pattern formation. }When
the injected field is a plane wave propagating along $z$, the amplitude $E$
is a constant that we take real and positive without loss of generality. In such a case Eqs. (\ref{model_norm}) admit spatially homogeneous solutions which coincide with the well-known ones of the single-mode models \cite{OM}, Eqs. (\ref{model_norm}) with $\nabla\rightarrow 0$.

For the linear model (LM), $\mu=1$, the stationary homogeneous solutions verify $\bar{Z}_{\mathrm{LM}}=|\bar{%
F}_{\mathrm{LM}}|^{2}$ with 
\begin{equation}
E^{2}=\left[ 1+\left( \Delta +\bar{Z}_{\mathrm{LM}}\right) ^{2}\right] \bar{Z%
}_{\mathrm{LM}}.  \label{S}
\end{equation}%
(Overbars denote steady homogenous solutions.) This is the usual state
equation of the single-mode optomechanical cavity (or of a Kerr cavity),
which shows a bistable region whenever $\Delta <-\sqrt{3}$, see Figs. \ref%
{Fig2}(a) and \ref{Fig2}(b).

Differently, in the case of the quadratic model (QM), $\mu=2$, there are two homogeneous steady states that bifurcate one into the other: the
trivial state, for which $\bar{Z}_{\mathrm{QM}}=0$ and $\bar{F}_{\mathrm{QM}%
}=E/\left( 1-\mathrm{i}\Delta \right) $, and the nontrivial one 
\begin{equation}
\bar{Z}_{\mathrm{QM}}^{2}=\sqrt{E^{2}-1}-\Delta ,\text{ }|\bar{F}_{\mathrm{QM%
}}|^{2}=1,  \label{QC}
\end{equation}%
which exists only for $E^{2}>\left( 1+\Delta ^{2}\right) $. Note that for $%
\Delta <0$ there is a region where the trivial and nontrivial solutions
coexist, see Figs. \ref{Fig2}(c) and \ref{Fig2}(d).

These spatially homogeneous steady states can suffer different instabilities
leading to the appearance of new solutions, such as self-pulsing or
space-dependent solutions (through Hopf or pattern-forming bifurcations,
respectively). Standard linear stability analysis techniques \cite%
{Routh-Hurwitz} allow us to derive the location of these bifurcations in the
parameter space \cite{SupMat}. However the general analysis is quite
involved as up to five parameters enter the models, and hence we will
give the details elsewhere, commenting here only on some general trends and
focusing on examples which show that pattern formation is possible.

Both models show a static pattern forming instability, as well as a Hopf
instability. The latter leads either to a pulsing homogeneous state (homogeneous
Hopf) or to a pulsing structure (pattern forming Hopf) depending on the
system parameters. Together with the homogeneous stationary solutions $\bar{Z%
}$, in Fig. \ref{Fig2} we have marked the typical location of these
instabilities, as well as the regions of $\bar{Z}$ that become unstable. In
very general terms, we can say that the existence of pattern forming
instabilities requires a rigidity $\rho $ below some maximum value which,
e.g., for the linear model is unity for most detuning values, and can be
even larger for large negative detunings. As for the Hopf instability, we
can also say that the smaller $\gamma $ or the larger $\Omega $ are, the
closer it is to the limit of existence of the nonzero solutions, in
particular tending to invade the whole domain of existence of the nontrivial
solution for small $\gamma $ in the quadratic model.

In order to study the type of patterns appearing in the unstable regions, we
have performed a numerical analysis of the model equations in both one (1D)
and two (2D) transverse dimensions using periodic boundary conditions, check \cite{SupMat} for details. Our
analysis has revealed periodic patterns (hexagonal in 2D) and localized structures, as well as
more complex spatio-temporal phenomena. Remarkably, stable localized
structures appear in the regions where there is coexistence between the stable
lower branch and the unstable upper branch solutions in the linear model (see the inset of Fig. \ref{Fig2}a), or between the stable trivial and unstable nontrivial solutions in the quadratic one. We have checked that these localized structures can be written, erased, and moved individually, so that they behave as true cavity solitons \cite{Sol}. An important difference between the two models is that the quadratic one has intrinsic phase bistability, since $Z\left( \mathbf{\bar{r}},\tau \right) \rightarrow -Z\left( \mathbf{\bar{r}},\tau \right) $ leaves the equations invariant, which leads to the existence of domain walls that appear between two adjacent regions occupied by solutions differing only in their sign (see the inset of Fig. \ref{Fig2}d).

\textbf{Physical implementation. }State-of-the-art optomechanical setups allow for the use of silicon nitride membranes both as intracavity elements dispersively coupled to the light contained in the resonator \cite{Thompson08,Zwickl08,Jayich08,Wilson09,Sankey10,Karuza12,Karuza13,Karuza13bis} or directly as end mirrors, hence sensitive to radiation pressure \cite{Kemiktarak12,Kemiktarak12b,Kemiktarak13}. In all these experiments the membrane is held by a fixed frame, so that the membrane cannot be displaced as a whole in the axial direction. Such systems are then described by our 2D model equations (\ref{dAdtaux}) and (\ref{ecQ}) with $\mathbb{L}=-v^2\nabla^2$ and fixed boundary conditions. We have performed extensive numerical simulations of this scenario obtaining no stable 2D patterns. This has lead us to the conclusion that $\Omega_{\mathrm{m}}\neq 0$ is a necessary condition for spatial instabilities to occur, since otherwise it is not possible to have a sufficiently homogeneous mechanical background to sustain such structures, in agreement with the intuition we provided when first introducing the differential operator as $\mathbb{L}=\Omega _{\mathrm{m}}^{2}-v^{2}\nabla _{\bot }^{2}$. While it is true that this condition excludes the possibility of studying the phenomena predicted by our 2D models with the membrane-based setups mentioned above, in the following we show that under conditions leading to the excitation of one or few mechanical modes in one transverse direction, an effective homogeneous mode appears in the orthogonal direction, allowing for the observation of the patterns predicted by our models in 1D.

In order to see this, let us consider a mechanical field of the form $Q(\mathbf{r},t)=\cos(\pi y/L_y)Q_{1\mathrm{D}}(x,t)$, where $L_{y}$ is the length of the membrane in the $y$ direction, in which only the fundamental mode is assumed to be excited for simplicity. The application of the differential operator $\mathbb{L}=-v^2\nabla^2$ onto it leads to $\mathbb{L}Q\equiv\Omega_\mathrm{m,eff}^2(1-\rho_\mathrm{eff}^2\partial_{\bar{x}}^2)Q$, with $\Omega_\mathrm{m,eff}=\pi v/L_y$ and $\rho_\mathrm{eff} =v/\Omega_\mathrm{m,eff}l_\mathrm{c}=L_y/\pi l_\mathrm{c}$, which is the 1D version of the analogous term in Eq. (\ref{dZ}), allowing for homogeneous excitations along the $x$ axis. Thus, we obtain an effective 1D rigidity parameter which is solely controlled by the length of the membrane in the $y$ direction (normalized to $l_\mathrm{c}$). A relevant question is whether current setups can reach the condition $\rho_\mathrm{eff}\lesssim 1$ which typically leads to pattern formation for a wide range of the rest of parameters, as explained above. Taking as in \cite{Karuza13bis} a laser at $\lambda_\mathrm{L}=1064\mathrm{nm}$ and a cavity of $L=10\mathrm{cm}$ but smaller finesse $\mathcal{F}=500$ (it is 50,000 in \cite{Karuza13bis}), we get a diffraction length $l_{\mathrm{c}}=\sqrt{\lambda_\mathrm{L} L\mathcal{F}}/2\pi\approx 1\mathrm{mm}$. Since commercial membranes have typical sizes on the millimeter scale, this shows that current experiments have indeed access to the desired regime.

We have performed numerical simulations that confirm this simple picture, using a quasi-planar injection of finite width (supergaussian) \cite{SupMat}. The conditions for exciting just one or few mechanical modes are naturally accomplished by choosing a membrane short enough along the $y$ axis for the frequencies of the normal modes to be well spaced, but long enough along the $x$ axis for the desired spatial structures to exist. In Figs. \ref{Fig2}(e) and \ref{Fig2}(f) we show results for both the linear and quadratic models, in particular showing a cavity soliton from the linear model and a domain wall from the quadratic one. The particular parameters used in the simulations can be checked in \cite{SupMat}, but they are not of real relevance since the structures are robust and easily found within the region of the parameter space where they are expected to appear according to the effective 1D model.

\textbf{Conclusions.} We have proposed an optomechanical system in which
pattern formation can be observed. In our proposal, the mechanical degrees
of freedom come from a membrane which can be locally deformed by its
interaction with light. Two different models appear in this scenario, for
which we have been able to locate their pattern forming instabilities. An
important conclusion of our numerical investigations is that the existence
of such structures requires the presence of a sufficiently homogeneous
mechanical mode, and we have proposed realistic implementations of the 1D
models based on currently available membranes, proving how patterns
(solitons and domain walls) indeed appear in such systems. Future venues
will include the study of the quantum spatial correlations present on (and
between) the optical and mechanical fields, which may lead to noncritical
squeezing and multipartite entanglement, similarly to what has been found in
optical parametric oscillators \cite{OPOs,NonCritical}.

\begin{acknowledgements}
We acknowledge fruitful discussions with Chiara Molinelli. This work has
been supported by the Spanish Government and the European Union FEDER
through Projects FIS2011-26960 and FIS2014-60715-P. J.R.-R. is a grant holder of the FPU program
of the Ministerio de Educaci\'{o}n, Cultura y Deporte (Spain). C.N.-B.
acknowledges the financial support of the Future and Emerging Technologies
(FET) programme within the Seventh Framework Programme for Research of the
European Commission, under the FET-Open grant agreement MALICIA, number
FP7-ICT-265522, and of the Alexander von Humboldt Foundation through its
Fellowship for Postdoctoral Researchers.
\end{acknowledgements}

\newpage \hspace{1cm} \newpage

\begin{center}
\textbf{Supplemental material}
\end{center}

In this supplemental material we offer a detailed derivation of the model equations, and give more details about the linear stability analysis performed onto the homogeneous stationary solutions present in the system, as well as about the numerical simulations performed both for the model and the proposed implementation.

\textbf{Derivation of the light field equation.} In this section we offer a
detailed derivation of Eq. (2) of the main text, which describes the
evolution of the optical field at the plane of the end mirror.

We will denote by $r_{1}$ ($r_{2}$) the reflection coefficients of the left
(right) cavity mirrors, $1-\left\vert r_{j}\right\vert ^{2}=T_{j}$ being the
corresponding transmittances, which are assumed very close to zero: good
cavity limit. The left (right) mirror is located at $z=0$ $\left( L\right) $%
, and we assume for definiteness that light is injected through the left
mirror.

In the absbence of illumination the membrane has an equilibrium position at $%
z=z_{0}$, while in the presence of optical fields any point $\mathbf{r}$ of
the membrane will be displaced along the cavity axis by $Q\left( \mathbf{r}%
,t\right)$ from equilibrium, which defines the mechanical field
introduced in the main text. We treat this deformable membrane as a thin,
lossless symmetric beam splitter, with (complex) transmission and reflection
coefficients denoted by $\tau _{\pm }$ and $\varrho _{\pm }$, where the
subscript refers to the side of the membrane from which the beam emerges after transmission or reflection ($+$ for right and $-$ for left). As in any lossless beam
splitter \cite{BS}, the phases of these coefficients satisfy%
\begin{equation}
\arg \left( \varrho _{+}\right) +\arg \left( \varrho _{-}\right) -\arg
\left( \tau _{+}\right) -\arg \left( \tau _{-}\right) =\pi ,  \label{BSrel}
\end{equation}%
while further for a symmetric one,%
\begin{equation}
\left\vert \tau _{\pm }\right\vert =\tau ,\ \ \ \left\vert \varrho _{\pm
}\right\vert =\varrho ,
\end{equation}%
with the relation $\tau ^{2}+\varrho ^{2}=1$.

\begin{figure}[t]
\includegraphics[width=0.75\columnwidth]{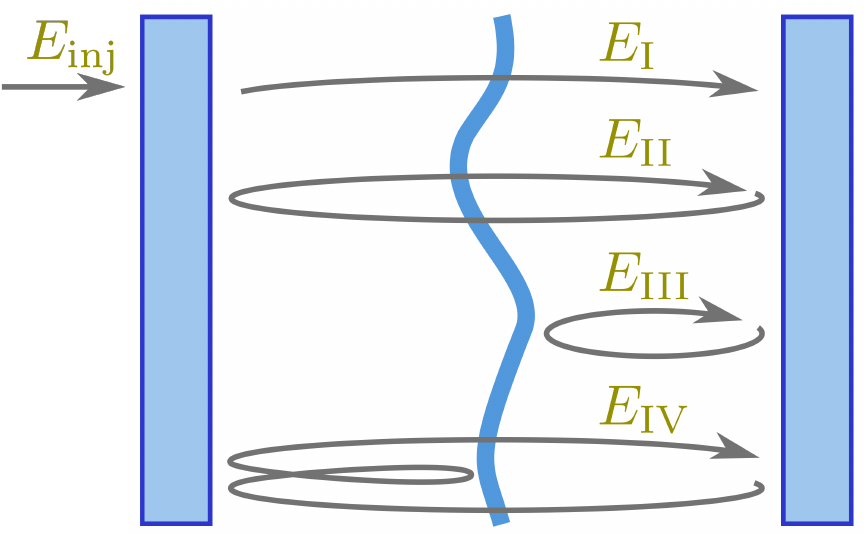}
\caption{Sketch of the paths travelled by the main waves superposing at the
right cavity mirror at a given instant.}
\label{FigModelDer}
\end{figure}

At any point in space and time, we write the optical field as $E\left( z,%
\mathbf{r},t\right) =E_{+}\left( z,\mathbf{r},t\right) +E_{-}\left( z,%
\mathbf{r},t\right) $, which is a sum of two waves traveling to the right ($%
+ $) and to the left ($-$). We choose arbitrarily to derive the evolution
equation for the field $E_{+}\left( L,\mathbf{r},t\right) $ impinging the
right cavity mirror. Such an equation is derived by following the usual
approach of propagating the field along the resonator (see e.g. \cite%
{hyperbolic}), assuming that any modification of the field along a cavity
roundtrip (due to diffraction and to transmission and reflection on the
membrane or on the cavity mirrors) is small. This means that we are
considering (i) short enough propagation distances (either geometrically
small, or optically small, as in quasi self-imaging resonators \cite{SI}),
and (ii) membranes with very small reflectivities, $\varrho \ll 1$. The
presence of the intracavity membrane makes the derivation a bit complicated
because, at any instant, $E_{+}\left( L,\mathbf{r},t\right) $ is the
superposition of infinitely many contributions, corresponding to waves that
have travelled in the cavity through paths with different combinations of
transmissions and reflections in the membrane and the mirrors, and meet at
the right mirror. However, as the membrane reflectivities $\varrho _{\pm }$
are assumed small, $E_{+}\left( L,\mathbf{r},t\right) $ can be approximated
at any instant as the sum of just four partial waves, as sketched in Figure %
\ref{FigModelDer}: (I) the injected field transmitted through the input
mirror and the membrane (call it $E_{\mathrm{I}}$), (II) the field that,
after reflection on the right cavity mirror, has performed a full cavity
roundtrip just by transmitting through the membrane (call it $E_{\mathrm{II}%
} $); (III) the field that, after reflection on the right cavity mirror,
relfects back from the right face of the membrane (call it $E_{\mathrm{III}}$%
); and (IV), the field that, after transmission through the membrane and
reflection on the left cavity mirror, has reflected from the left side of
the membrane, reflected again from the left mirror, and finally transmitted
through the membrane (call it $E_{\mathrm{IV}}$). Any other partial wave has
an amplitude on the order of $\varrho ^{2}$ or smaller, which we neglect.
Hence we write%
\begin{equation}
E_{+}\left( L,\mathbf{r},t\right) =E_{\mathrm{I}}\left( \mathbf{r},t\right)
+E_{\mathrm{II}}\left( \mathbf{r},t\right) +E_{\mathrm{III}}\left( \mathbf{r}%
,t\right) +E_{\mathrm{IV}}\left( \mathbf{r},t\right) ,  \label{E+total}
\end{equation}%
where the four partial waves can be written as 
\begin{subequations}
\label{partialwaves}
\begin{eqnarray}
E_{\mathrm{I}}\left( \mathbf{r},t\right) &=&K_{\mathrm{I}}\mathcal{U}_{L}E_{%
\mathrm{inj}}\left( 0,\mathbf{r},t-t_{\mathrm{c}}/2\right) , \\
E_{\mathrm{II}}\left( \mathbf{r},t\right) &=&K_{\mathrm{II}}\mathcal{U}%
_{2L}E_{+}\left( L,\mathbf{r},t-t_{\mathrm{c}}\right) , \\
E_{\mathrm{III}}\left( \mathbf{r},t\right) &=&K_{\mathrm{III}}\mathcal{U}%
_{L_{2}}e^{-2\mathrm{i}k_{\mathrm{L}}Q\left( \mathbf{r},t-t_{2}\right) } \\
&&\times \mathcal{U}_{L_{2}}E_{+}\left( L,\mathbf{r},t-2t_{2}\right) , 
\notag \\
E_{\mathrm{IV}}\left( \mathbf{r},t\right) &=&K_{\mathrm{IV}}\mathcal{U}%
_{L+L_{1}}e^{2\mathrm{i}k_{\mathrm{L}}Q\left( \mathbf{r},t-t_{\mathrm{c}%
}/2-t_{1}\right) } \\
&&\times \mathcal{U}_{L+L_{1}}E_{+}\left( L,\mathbf{r},t-t_{\mathrm{c}%
}-2t_{1}\right) ,  \notag
\end{eqnarray}%
with 
\end{subequations}
\begin{eqnarray}
K_{\mathrm{I}} &=&\sqrt{T_{1}}\tau _{+}, \\
K_{\mathrm{II}} &=&r_{2}r_{1}\tau _{+}\tau _{-}, \\
K_{\mathrm{III}} &=&r_{2}\rho _{+}, \\
K_{\mathrm{IV}} &=&r_{1}^{2}r_{2}\tau _{+}\tau _{-}\varrho _{-}.
\end{eqnarray}%
The operator $\mathcal{U}_{d}=\exp \left[ \mathrm{i}(d/2k_{\mathrm{L}%
})\nabla _{\bot }^{2}\right] $ accounts for diffraction in the paraxial
approximation, corresponding to a propagation distance equal to $d$. Here $%
L_{1}=z_{0}$ and $L_{2}=L-z_{0}$, $t_{1,2}=L_{1,2}/c$, and $t_{\mathrm{c}%
}=2\left( t_{1}+t_{2}\right) =2L/c$ is the cavity roundtrip time. The
factors $e^{\pm 2\mathrm{i}k_{\mathrm{L}}Q\left( \mathbf{r},t\right) }$
model the phase front modification produced by the reflection on the
membrane in the paraxial approximation.

Before continuing it proves useful to express the coefficients $K_{\mathrm{II%
}}$, $K_{\mathrm{III}}$ and $K_{\mathrm{IV}}$ in terms of the modulus and
argument of $K_{\mathrm{II}}$. Let us then write 
\begin{subequations}
\begin{align}
K_{\mathrm{II}}& =r\exp \left( \mathrm{i}\theta \right) , \\
K_{\mathrm{III}}& =K_{+}r\exp \left( \mathrm{i}\theta \right) , \\
\ K_{\mathrm{IV}}& =K_{-}r\exp \left( \mathrm{i}\theta \right) ,
\end{align}%
with 
\end{subequations}
\begin{subequations}
\begin{align}
r& =\left\vert r_{1}r_{2}\tau _{+}\tau _{-}\right\vert , \\
\theta & =\arg \left( r_{1}\right) +\arg \left( r_{2}\right) +\arg \left(
\tau _{+}\right) +\arg \left( \tau _{-}\right) , \\
K_{+}& =\frac{\varrho _{+}}{r_{1}\tau _{+}\tau _{-}},  \label{Kplus} \\
K_{-}& =r_{1}\varrho _{-}.  \label{Kminus}
\end{align}%
Remembering that the mirrors' transmittances are small, so we assume in
particular $T_{j}=O\left( \varrho \right) $, we have $r=\left\vert
r_{1}r_{2}\right\vert \tau ^{2}=\left\vert r_{1}r_{2}\right\vert +O\left(
\varrho ^{2}\right) $. Expressing $\left\vert r_{j}\right\vert =\sqrt{1-T_{j}%
}$, we get then 
\end{subequations}
\begin{equation}
r=1-\frac{1}{2}\left( T_{1}+T_{2}\right) +O\left( \varrho ^{2}\right) .
\end{equation}%
Further, taking into account that $\arg \left( K_{+}\right) +\arg \left(
K_{-}\right) =\pi \,$, see (\ref{Kplus}), (\ref{Kminus}), and (\ref{BSrel}),
we note that%
\begin{equation}
K_{\pm }=\mp \varrho \exp \left( \mp \mathrm{i}\beta \right) +O(\varrho
^{2}),
\end{equation}%
where we have defined $\beta =\arg \left( r_{1}\right) +\arg \left( \varrho
_{-}\right) $.

We will denote the slowly varying complex amplitude of $E_{+}\left( L,%
\mathbf{r},t\right) $ by $A\left( \mathbf{r},t\right) $, so that $%
E_{+}\left( L,\mathbf{r},t\right) =\mathrm{i}\mathcal{V}A\left( \mathbf{r}%
,t\right) e^{\mathrm{i}k_{\mathrm{L}}L-\mathrm{i}\omega _{\mathrm{L}}t}$,
where $\mathcal{V}=\sqrt{\hbar \omega _{\mathrm{c}}/4\varepsilon _{0}L}$ is
a suitable factor with dimensions of voltage, chosen in such a way that $%
|A\left( \mathbf{r},t\right) |^{2}$ can be interpreted as the number of
photons per unit area which arrive at point $\mathbf{r}$ of the mirror
during a roundtrip, what is convenient to make contact with quantum theory,
see the next section. When the expressions (\ref{partialwaves}) of the
partial waves are introduced into (\ref{E+total}) and all the fields are
expressed in terms of the complex amplitude $A$, we get, shifting time for
convenience as $t\rightarrow t+t_{\mathrm{c}}$,%
\begin{align}
& A\left( \mathbf{r},t+t_{\mathrm{c}}\right) -A\left( \mathbf{r},t\right) =%
\sqrt{T_{1}}\tau _{-}\mathcal{U}_{L}A_{\mathrm{inj}}\left( 0,\mathbf{r},t+t_{%
\mathrm{c}}/2\right)  \notag \\
& +\left( re^{\mathrm{i}\Psi }\mathcal{U}_{2L}-1\right) A\left( \mathbf{r}%
,t\right)  \label{difA} \\
& -\varrho e^{\mathrm{i}\Psi }r\mathcal{U}_{L_{2}}e^{-2\mathrm{i}k_{\mathrm{L%
}}\left[ \tilde{z}_{0}+Q\left( \mathbf{r},t+t_{\mathrm{c}}-t_{2}\right) %
\right] }\mathcal{U}_{L_{2}}A\left( \mathbf{r},t+2t_{1}\right)  \notag \\
& +\varrho e^{\mathrm{i}\Psi }r\mathcal{U}_{L+L_{1}}e^{2\mathrm{i}k_{\mathrm{%
L}}\left[ \tilde{z}_{0}+Q\left( \mathbf{r},t+t_{\mathrm{c}}+t_{2}\right) %
\right] }\mathcal{U}_{L+L_{1}}A\left( \mathbf{r},t-2t_{1}\right) ,  \notag
\end{align}%
where we have subtracted $A\left( \mathbf{r},t\right) $ from both sides for
convenience, and we have defined 
\begin{subequations}
\begin{eqnarray}
\Psi &=&2k_{\mathrm{L}}L+\theta ,  \label{Psiq0} \\
\tilde{z}_{0} &=&z_{0}+\beta /2k_{\mathrm{L}}.
\end{eqnarray}

Note that the left hand side of Eq. (\ref{difA}) can be approximated by $t_{%
\mathrm{c}}\partial _{t}A\left( \mathbf{r},t\right) $, whenever its right
hand side is small; more rigorously if it is of the form $\mathcal{L}A\left( 
\mathbf{r},t\right) $, with $\mathcal{L}$ a small operator. Inspection of
the equation shows that this is satisfied provided that $\varrho \ll 1$, as
already assumed, and $e^{\mathrm{i}\Psi }r\mathcal{U}_{2L}-1=O\left( \varrho
\right) $. This second condition requires (i) $\Psi =2m\pi +\delta $, with $%
m\in 
%TCIMACRO{\U{2124} }%
%BeginExpansion
\mathbb{Z}
%EndExpansion
$ and $\delta $ an $O\left( \varrho \right) $ normalized detuning whose
value is controled by the injection frequency, see Eq. (\ref{Psiq0}), and
(ii) $\mathcal{U}_{2L}$ can be approximated as $1+\mathrm{i}\left( L/k_{%
\mathrm{L}}\right) \nabla _{\bot }^{2}$, with the effect of the last term on
the order of $\varrho $, which is effected by the choice of a sufficiently
small value of $L$ (small diffraction). Under these conditions, $e^{\mathrm{i%
}\Psi }$, $r$, $\mathcal{U}_{L_{2}}$, and $\mathcal{U}_{L+L_{1}}$ in the
last two terms can be approximated by $1$ (note that these terms already
contain $\varrho $ as a factor), while $A\left( \mathbf{r},t\pm
2t_{1}\right) $ can be set to $A\left( \mathbf{r},t\right) $, since $A\left( 
\mathbf{r},t\pm 2t_{1}\right) \simeq A\left( \mathbf{r},t\right) \pm
2t_{1}\partial _{t}A\left( \mathbf{r},t\right) $, but $2t_{1}\partial
_{t}A\left( \mathbf{r},t\right) \leq t_{\mathrm{c}}\partial _{t}A\left( 
\mathbf{r},t\right) =O(\varrho )$. Accordingly Eq. (\ref{difA}) can be
approximated as 
\end{subequations}
\begin{align}
& t_{\mathrm{c}}\partial _{t}A\left( \mathbf{r},t\right) =A_{0}\left( 
\mathbf{r},t\right) -\frac{T}{2}A+\mathrm{i}\left( \delta +\frac{L}{k_{%
\mathrm{L}}}\nabla _{\bot }^{2}\right) A \\
& +\varrho \left[ e^{2\mathrm{i}k_{\mathrm{L}}\left[ \tilde{z}_{0}+Q\left( 
\mathbf{r},t+t_{\mathrm{c}}+t_{2}\right) \right] }-e^{-2\mathrm{i}k_{\mathrm{%
L}}\left[ \tilde{z}_{0}+Q\left( \mathbf{r},t+t_{\mathrm{c}}-t_{2}\right) %
\right] }\right] A,  \notag
\end{align}%
where $A_{0}\left( \mathbf{r},t\right) =\tau _{-}\sqrt{T_{1}}\mathcal{U}%
_{L}A_{\mathrm{inj}}\left( 0,\mathbf{r},t+t_{\mathrm{c}}/2\right) $ is the
injected field at the plane of the right mirror. Note that $A_{0}\left( 
\mathbf{r},t\right) \approx \sqrt{T_{1}}e^{\mathrm{i}\arg \left( \tau
_{-}\right) }A_{\mathrm{inj}}\left( 0,\mathbf{r},t+t_{\mathrm{c}}/2\right) $%
. There remains making a last simplification, consisting in approximating $%
Q\left( \mathbf{r},t+t_{\mathrm{c}}\pm t_{2}\right) $ by $Q\left( \mathbf{r}%
,t\right) $. Note that $Q\left( \mathbf{r},t+t_{\mathrm{c}}\pm t_{2}\right)
\approx Q\left( \mathbf{r},t\right) +\left( t_{c}\pm t_{2}\right) \partial
_{t}Q\left( \mathbf{r},t\right) $, and we have checked self-consistently
that the last term is on the order of $\varrho $ or smaller in our
simulations. This last approximation leads us then to%
\begin{eqnarray}
\partial _{t}A\left( \mathbf{r},t\right) &=&\frac{1}{t_{\mathrm{c}}}%
A_{0}\left( \mathbf{r},t\right) \mathcal{-}\frac{T}{2t_{\mathrm{c}}}A+%
\mathrm{i}\left( \frac{\delta }{t_{\mathrm{c}}}+\frac{L}{k_{\mathrm{L}}t_{%
\mathrm{c}}}\nabla _{\bot }^{2}\right) A  \notag \\
&&+\mathrm{i}\frac{2\varrho }{t_{\mathrm{c}}}\sin \left[ 2k_{\mathrm{L}%
}\left( \tilde{z}_{0}+Q\right) \right] A,  \label{dAdt_gen}
\end{eqnarray}%
which coincides with Eq. (\ref{dAdtaux}) of the main text once we introduced
the parameters defined there, and the normalized detuning apearing on it is
identified with $2\delta /T$. In addition, note that in the main text we
have chosen $\beta =-\pi /2$ and $\arg \left( \tau _{-}\right) =0$ to
simplify the notation and make connection with previous works.

\textbf{Derivation of the coupling in the mechanical equation.} In this
section we derive the optomechanical coupling term appearing on the right
hand side of the mechanical equation of motion, Eq. (\ref{ecQ}) in the main
text. We show in the following that this is easily done from the coupling
term derived for the optical field in the previous section, together with a
(momentary) quantum description of the system. In such description, the
fields $A$ and $Q$ are replaced by two operators $\hat{A}$ and $\hat{Q}$,
obeying standard equal-time commutation relations $\left[ \hat{A}\left( 
\mathbf{r},t\right) ,\hat{A}^{\dag }\left( \mathbf{r}^{\prime },t\right) %
\right] =\delta ^{2}\left( \mathbf{r}-\mathbf{r}^{\prime }\right) $ \cite%
{OPOs} and $\left[ \hat{Q}\left( \mathbf{r},t\right) ,\hat{P}\left( \mathbf{r%
}^{\prime },t\right) \right] =\mathrm{i}\hbar \delta ^{2}\left( \mathbf{r}-%
\mathbf{r}^{\prime }\right) $, with $\hat{P}=\sigma \partial _{t}\hat{Q}$
the momentum density field of the membrane ($\sigma $ is its mass surface
density). The coupling between the optical and mechanical fields is
described in the quantum theory through an interaction Hamiltonian $\hat{H}_{%
\mathrm{int}}$, which contributes to the equation of motion of any operator $%
\hat{O}$ according to Heisenberg's equation%
\begin{equation}
\partial _{t}\hat{O}|_{\mathrm{int}}=\frac{\mathrm{i}}{\hbar }\left[ \hat{H}%
_{\mathrm{int}},\hat{O}\right] .  \label{HeisEq}
\end{equation}%
Now, from the optical equation that we derived in the previous section, we
know that when Eq. (\ref{HeisEq}) is applied to the field $\hat{A}\left( 
\mathbf{r},t\right) $, we should get 
\begin{equation}
\left. \partial _{t}\hat{A}\left( \mathbf{r},t\right) \right\vert _{\mathrm{%
int}}=\mathrm{i}\frac{2\varrho }{t_{\mathrm{c}}}\sin \left\{ 2k_{\mathrm{L}}%
\left[ \tilde{z}_{0}+\hat{Q}\left( \mathbf{r},t\right) \right] \right\} \hat{%
A},
\end{equation}%
and hence the interaction Hamiltonian must have the form%
\begin{eqnarray}
\hat{H}_{\mathrm{int}} &=&-\hbar \int d^{2}\mathbf{r}\frac{2\varrho }{t_{%
\mathrm{c}}}\sin \left\{ 2k_{\mathrm{L}}\left[ \tilde{z}_{0}+\hat{Q}\left( 
\mathbf{r},t\right) \right] \right\} \\
&&\text{ \ \ \ \ \ \ \ \ \ \ \ \ \ \ \ \ }\times \hat{A}^{\dag }\left( 
\mathbf{r},t\right) \hat{A}\left( \mathbf{r},t\right) .  \notag
\end{eqnarray}%
On the other hand, once we know the interaction Hamiltonian, we can
particularize the Heisenberg equation (\ref{HeisEq}) to the mechanical
momentum density field, obtaining 
\begin{eqnarray}
\left. \partial _{t}\hat{P}\left( \mathbf{r},t\right) \right\vert _{\mathrm{%
int}} &=&\frac{4k_{\mathrm{L}}\varrho }{t_{\mathrm{c}}}\cos \left\{ 2k_{%
\mathrm{L}}\left[ \tilde{z}_{0}+\hat{Q}\left( \mathbf{r},t\right) \right]
\right\} \\
&&\text{ \ \ \ \ \ \ \ \ \ }\times \hat{A}^{\dag }\left( \mathbf{r},t\right) 
\hat{A}\left( \mathbf{r},t\right) ,  \notag
\end{eqnarray}%
where we have used the property%
\begin{equation}
\left[ F\left( \hat{Q}\left( \mathbf{r}^{\prime },t\right) \right) ,\hat{P}%
\left( \mathbf{r},t\right) \right] =\mathrm{i}\hbar \left( \partial
F/\partial \hat{Q}\right) \delta ^{2}\left( \mathbf{r}-\mathbf{r}^{\prime
}\right) ,
\end{equation}%
valid for any function $F$; the classical limit of this expression, $\{\hat{Q%
},\hat{P},\hat{A},\hat{A}^{\dagger }\}\rightarrow \{Q,\sigma \partial
_{t}Q,A,A^{\ast }\}$, provides the coupling appearing in Eq. (\ref{ecQ}) of
the main text, 
\begin{equation}
\partial _{t}^{2}Q\left( \mathbf{r},t\right) |_{\mathrm{int}}=\frac{4\hbar
k_{\mathrm{L}}\rho }{\sigma t_{\mathrm{c}}}\cos \left\{ 2k_{\mathrm{L}}\left[
\tilde{z}_{0}+Q\left( \mathbf{r},t\right) \right] \right\} |A\left( \mathbf{r%
},t\right) |^{2}\mathrm{,}
\end{equation}%
where note that in main text we have taken $\beta =-\pi /2$ and hence $2k_{%
\mathrm{L}}\tilde{z}_{0}=2k_{\mathrm{L}}z_{0}-\pi /2$ for simplicity and
definitiness.

\textbf{Linear and quadratic models and their normalization.} Starting from
the general model equations (\ref{dAdtaux}) and (\ref{ecQ}) of the main text, in this section we provide a detailed derivation of the quadratic and linear models
introduced in Eqs. (\ref{model_norm}). Consider the
interaction terms of Eqs. (\ref{dAdtaux}) and (\ref{ecQ}), which can be
written as 
\begin{subequations}
\begin{eqnarray}
\partial _{t}A|_{\mathrm{int}} &=&-\mathrm{i}\frac{4\gamma _{\mathrm{c}%
}\varrho }{T}\cos \left[ 2k_{\mathrm{L}}\left( z_{0}+Q\right) \right] A, \\
\partial _{t}^{2}Q|_{\mathrm{int}} &=&\frac{4\varrho \hbar k_{\mathrm{L}}}{%
\sigma t_{\mathrm{c}}}\sin \left[ 2k_{\mathrm{L}}\left( z_{0}+Q\right) %
\right] \left\vert A\right\vert ^{2}.
\end{eqnarray}
\end{subequations}
Taking into account that the mechanical displacement is typically much
smaller than the optical wavelength, we can expand the trigonometric
functions around $Q=0$, obtaining
\begin{subequations}
\begin{align}
\cos \left[ 2k_{\mathrm{L}}\left( z_{0}+Q\right) \right]  &\approx \cos
(2k_{\mathrm{L}}z_{0})(1-k_{\mathrm{L}}^{2}Q^{2})
\\
&\hspace{0.6cm}-2k_{\mathrm{L}}\sin \left(2k_{\mathrm{L}}z_{0}\right) Q,\nonumber
\\
\sin \left[ 2k_{\mathrm{L}}\left( z_{0}+Q\right) \right]  &\approx \sin
\left( 2k_{\mathrm{L}}z_{0}\right) (1-k_{\mathrm{L}}^{2}Q^{2})
\\
&\hspace{0.6cm}+2k_{\mathrm{L}}\cos \left( 2k_{\mathrm{L}}z_{0}\right) Q.\nonumber
\end{align}
\end{subequations}
The interaction terms admit then different approximations depending on where
the membrane is located. In particular, when $\cos (2k_{\mathrm{L}%
}z_{0})=\pm 1$ and $\sin (2k_{\mathrm{L}}z_{0})=0$, that is, $z_{0}/\lambda
_{\mathrm{L}}=n/4$ with $n=0,1,2,...$, the sign of the cosine being positive
(negative) for even (odd) $n$, we get
\begin{subequations}
\begin{eqnarray}
\partial _{t}A|_{\mathrm{int}} &\approx &\pm \mathrm{i}\frac{4\gamma _{%
\mathrm{c}}\varrho }{T}(k_{\mathrm{L}}^{2}Q^{2}-1)A, \\
\partial _{t}^{2}Q|_{\mathrm{int}} &\approx &\pm \frac{8\varrho \hbar k_{%
\mathrm{L}}^{2}}{\sigma t_{\mathrm{c}}}Q\left\vert A\right\vert ^{2},
\end{eqnarray}%
leading to the model equations
\end{subequations}
\begin{subequations}
\begin{align}
\partial _{t}A &=\gamma _{\mathrm{c}}\mathcal{E}+\gamma _{\mathrm{c}}\left( 
\mathcal{-}1+\mathrm{i}\Delta _{\pm }+\mathrm{i}l_{\mathrm{c}}^{2}\nabla
_{\bot }^{2}\right) A \\
&\hspace{2.5cm}\pm \mathrm{i}\frac{4\gamma _{\mathrm{c}}k_{\mathrm{L}}^{2}\varrho }{T}%
Q^{2}A,  \notag \\
\partial _{t}^{2}Q&+\gamma _{\mathrm{m}}\partial _{t}Q +(\Omega _{\mathrm{m}%
}^{2}-v^{2}\nabla _{\bot }^{2})Q \\
&\hspace{2.5cm}=\pm \frac{8\varrho \hbar k_{\mathrm{L}}^{2}}{\sigma t_{\mathrm{c}}}%
Q\left\vert A\right\vert ^{2},  \notag
\end{align}%
where $\Delta _{\pm }=\Delta \mp 4\varrho /T$ is a shifted detuning. It is
convenient both for the numerical and analytical analysis to introduce the
nomalized time $\tau =\gamma _{\mathrm{c}}t$ and spatial coordinates $%
\mathbf{\bar{r}}=\mathbf{r}/l_{\mathrm{c}}$, plus variables and parameters
\end{subequations}
\begin{eqnarray*}
Z &=&\sqrt{\frac{8\varrho }{T}}k_{\mathrm{L}}Q\text{, \ \ }F=\sqrt{\frac{%
8\varrho \hbar k_{\mathrm{L}}^{2}}{\sigma t_{\mathrm{c}}\Omega _{\mathrm{m}%
}^{2}}}A\text{,} \\
\gamma  &=&\frac{\gamma _{\mathrm{m}}}{\gamma _{\mathrm{c}}}\text{, \ }%
\Omega =\frac{\Omega _{\mathrm{m}}}{\gamma _{\mathrm{c}}}\text{, \ }E=\sqrt{%
\frac{8\varrho \hbar k_{\mathrm{L}}^{2}}{\sigma t_{\mathrm{c}}\Omega _{%
\mathrm{m}}^{2}}}\mathcal{E},
\end{eqnarray*}%
which transform the previous equations into
\begin{subequations}
\begin{gather}
\partial _{\tau }F=\left[ -1+\mathrm{i}\left( \Delta +\nabla ^{2}\pm
Z^{2}\right) \right] F+E, \\
\partial _{\tau }^{2}Z+\gamma \partial _{\tau }Z+\Omega ^{2}\left( 1-\rho
^{2}\nabla ^{2}\right) Z=\pm \Omega ^{2}Z\left\vert F\right\vert ^{2},
\end{gather}%
where $\nabla ^{2}=l_{\mathrm{c}}^{2}\left( \partial _{x}^{2}+\partial
_{y}^{2}\right) =\left( \partial _{\bar{x}}^{2}+\partial _{\bar{y}%
}^{2}\right) $, and $\rho =v/\Omega _{\mathrm{m}}l_{\mathrm{c}}$ is the
`effective rigidity parameter' defined in the main text. In the case of the
positive sign, this is the equation that we introduced in the main text
defining the quadratic model, Eqs. (\ref{model_norm}) with $\mu =2$. The
negative sign case is not interesting because no pattern forming
instabilities are found, and hence we have not introduced it in the main
text.

Let us consider now the case $\cos (2k_{\mathrm{L}}z_{0})=0$ and $\sin (2k_{%
\mathrm{L}}z_{0})=\pm 1$, that is, $z_{0}/\lambda _{\mathrm{L}}=(2n+1)/8$
with $n=0,1,2,...$, the sign of the sine function being positive (negative)
for even (odd) $n$. This choice leads to
\end{subequations}
\begin{subequations}
\begin{eqnarray}
\partial _{t}A|_{\mathrm{int}} &\approx &\pm \mathrm{i}\frac{8\gamma _{%
\mathrm{c}}k_{\mathrm{L}}\varrho }{T}QA, \\
\partial _{t}^{2}Q|_{\mathrm{int}} &\approx &\pm \frac{4\varrho \hbar k_{%
\mathrm{L}}}{\sigma t_{\mathrm{c}}}\left\vert A\right\vert ^{2},
\end{eqnarray}%
and hence to the model equations
\end{subequations}
\begin{subequations}
\begin{align}
\partial _{t}A &=\gamma _{\mathrm{c}}\mathcal{E}+\gamma _{\mathrm{c}}\left( 
\mathcal{-}1+\mathrm{i}\Delta +\mathrm{i}l_{\mathrm{c}}^{2}\nabla _{\bot
}^{2}\right) A \\
&\hspace{2.5cm}\pm \mathrm{i}\frac{8\gamma _{\mathrm{c}}k_{\mathrm{L}}\varrho }{T}QA, 
\notag \\
\partial _{t}^{2}Q&+\gamma _{\mathrm{m}}\partial _{t}Q +(\Omega _{\mathrm{m}%
}^{2}-v^{2}\nabla _{\bot }^{2})Q \\
&\hspace{2.5cm}=\pm \frac{4\varrho \hbar k_{\mathrm{L}}}{\sigma t_{\mathrm{c}}}\left\vert
A\right\vert ^{2}.  \notag
\end{align}%
In this case, the most convenient normalization is the same as before,
except for the fields and pump parameter, which are normalized as 
\end{subequations}
\begin{eqnarray}
Z &=&\pm \frac{8k_{\mathrm{L}}\varrho }{T}Q\text{, \ \ }F=\sqrt{\frac{%
32\varrho ^{2}\hbar k_{\mathrm{L}}^{2}}{\sigma t_{\mathrm{c}}\Omega _{%
\mathrm{m}}^{2}}}A\text{,} \\
E &=&\sqrt{\frac{32\varrho ^{2}\hbar k_{\mathrm{L}}^{2}}{\sigma t_{\mathrm{c}%
}\Omega _{\mathrm{m}}^{2}}}\mathcal{E}. \nonumber
\end{eqnarray}%
These normalizations transform the previous equations into
\begin{subequations}
\begin{gather}
\partial _{\tau }F=\left[ -1+\mathrm{i}\left( \Delta +\nabla ^{2}+Z\right) %
\right] F+E, \\
\partial _{\tau }^{2}Z+\gamma \partial _{\tau }Z+\Omega ^{2}\left( 1-\rho
^{2}\nabla ^{2}\right) Z=\Omega ^{2}\left\vert F\right\vert ^{2},
\end{gather}%
precisely the equation that we introduced in the main text defining the
linear model, Eqs. (\ref{model_norm}) with $\mu =1$.

\textbf{Details of the linear stability analysis. }In the following we
explain how we have performed the stability analysis of the homogeneous,
stationary solutions associated to the model equations (\ref{model_norm}),
which we gave in the main text. We have followed the standard linear
stability analysis, by studying the evolution of small perturbations $\left(
\delta F,\delta Z\right) $ added to the steady solution $\left( \bar{F},\bar{%
Z}\right) $. Upon linearizing the model equations (\ref{model_norm}) with
respect to $\left( \delta F,\delta Z\right) $, expressing them in terms of
the normal mode basis of the uncoupled mechanical and optical systems (plane
waves) as $\delta F(\mathbf{\bar{r}},\tau )=\sum_{\mathbf{k}}\phi _{\mathbf{k%
}}\left( \tau \right) e^{\mathrm{i}\mathbf{k}\cdot \mathbf{\bar{r}}} $ and $%
\delta Z(\mathbf{\bar{r}},\tau )=\sum_{\mathbf{k}}\zeta _{\mathbf{k}}\left(
\tau \right) e^{\mathrm{i}\mathbf{k}\cdot \mathbf{\bar{r}}}$ (note that $%
\zeta _{-\mathbf{k}}^{\ast }=\zeta _{\mathbf{k}}$ because $\delta Z$ is a
real field), and equating coefficients of like exponentials, we get, for
each $\mathbf{k}$, a linear system of differential equations for the modal
perturbations $\mathbf{v}\equiv \left( \phi _{\mathbf{k}},\phi _{-\mathbf{k}%
}^{\ast },\zeta _{\mathbf{k}}\right) $. Owed to this linearity and the time
invariance of the system, its solutions have the form $\mathbf{v}\left( \tau
\right) =\mathbf{v}\left( 0\right) e^{\lambda \tau }$. Upon making such
substitution, a homogeneous linear system of algebraic equations in $\mathbf{%
v}\left( 0\right) $ is got, whose condition for existence of nontrivial
solutions can be written as $C\left( k^{2};\lambda \right) \equiv
\sum_{n=0}^{4}c_{n}\left( k^{2}\right) \lambda ^{n}=0$, where $k=|\mathbf{k|}
$.

In the case of the linear model and after simple algebra, we get 
\end{subequations}
\begin{subequations}
\begin{align}
c_{4}& =1,\text{ \ \ }c_{3}=2+\gamma , \\
c_{2}& =1+2\gamma +\Omega _{k}^{2}+\Delta _{k}^{2}, \\
c_{1}& =\gamma (1+\Delta _{k}^{2})+2\Omega _{k}^{2}, \\
c_{0}& =2\bar{Z}_{\mathrm{LM}}\Delta _{k}\Omega ^{2}+(1+\Delta
_{k}^{2})\Omega _{k}^{2},
\end{align}%
where $\Omega _{k}^{2}\equiv \Omega ^{2}\left( 1+\rho ^{2}k^{2}\right) $ and 
$\Delta _{k}\equiv \Delta -k^{2}+\bar{Z}_{\mathrm{LM}}$. On the other hand,
the quadratic model leads to 
\end{subequations}
\begin{subequations}
\begin{eqnarray}
c_{4} &=&c_{3}=0\text{, \ \ }c_{2}=1\text{, \ \ }c_{1}=\gamma , \\
c_{0} &=&\Omega ^{2}[1+\rho ^{2}k^{2}-E^{2}/(1+\Delta ^{2})],
\end{eqnarray}%
for the trivial solution ($\bar{Z}_{\mathrm{QM}}=0$), and 
\end{subequations}
\begin{subequations}
\begin{align}
c_{4}& =1,\text{ \ \ }c_{3}=2+\gamma , \\
c_{2}& =1+2\gamma +\mu _{k}^{2}+\Omega ^{2}\rho ^{2}k^{2}, \\
c_{1}& =\gamma (1+\mu _{k}^{2})+2\Omega ^{2}\rho ^{2}k^{2}, \\
c_{0}& =[\rho ^{2}k^{2}(1+\mu _{k}^{2})+4\mu _{k}(\bar{Z}_{\mathrm{QM}%
}^{2}-\Delta )]\Omega ^{2},
\end{align}
\end{subequations}
for the nontrivial solution $\bar{Z}_{\mathrm{QM}}^{2}=\sqrt{E^{2}-1}$, with 
$\mu _{k}=$ $\bar{Z}_{\mathrm{QM}}^{2}-k^{2}$.

\begin{figure*}[t]
\centering
\includegraphics[width=\textwidth]{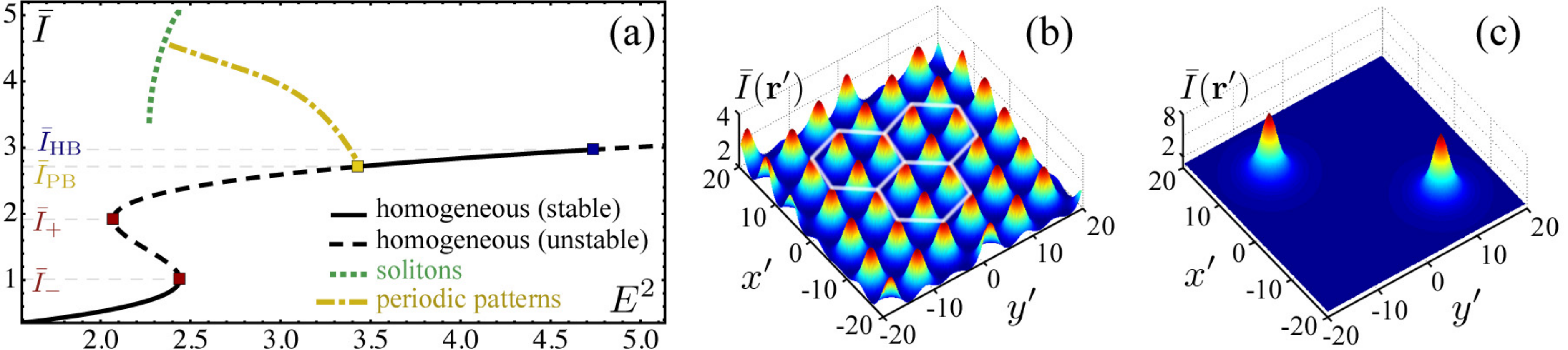}
\caption{Numerical simulation of the linear model equations. In (a) we show
the bifurcation diagram for $\Delta =-2.2$, $\protect\gamma =0.1$, $\Omega
=10$, and $\protect\rho =1.13$, parameters which we use in this example. The
intensity $\bar{I}$ of the homogeneous steady-state solution is plotted as a
function of the injection $E^{2}$, denoting by solid (dashed) lines the
stable (unstable) regions. We mark with a yellow square the static
pattern-forming instability at $\bar{I}=\bar{I}_{\mathrm{PB}}$, as well as
the rest of instabilities mentioned in the text. Patterns are expected to
appear for injections within the unstable domain of the upper branch
occurring for $\bar{I}_{+}<\bar{I}<\bar{I}_{\mathrm{PB}}$. We have found two
types of patterms: periodic (hexagonal in 2D) patterns (b), whose maximum
(as obtained from a 1D simulation) is represented as a yellow dashed-dotted
line in their domain of existence; and localized structures (c), represented
as a green, dotted line, which can be ``written'' and ``erased''
individually at any desired position in the transverse plane by injecting an
pulse of Gaussian tranverse profile with the proper width and centered at
the corresponding point. At $\bar{I}=\bar{I}_{\mathrm{HB}}$ the steady state
undergoes a Hopf bifurcation leading to time-dependent long-time term
solutions (limit cycles) not shown in the figure.}
\label{Fig2D}
\end{figure*}

We observe that the growth exponents $\lambda (k^{2})$, solutions to $%
C\left( k^{2};\lambda \right) =0$, depend on $k$ and not on $\mathbf{k}$
because of the rotational invariance of both the steady state and the model
equations. Whenever $\mathrm{Re}\{\lambda \}<0$ for all $k$ the steady state
is stable, while if $\mathrm{Re}\{\lambda \}>0$ for some $k$ it is unstable.
The condition $\mathrm{Re}\{\lambda \}=0$ thus defines a possible
instability, or bifurcation, which is met either when $\lambda =0$ (static
or pitchfork instability: $c_{0}=0$) or when $\lambda =\mathrm{i}\sqrt{%
c_{1}/c_{3}}$ (self-pulsing or Hopf instability: $%
c_{1}c_{2}c_{3}=c_{4}c_{1}^{2}+c_{3}^{2}c_{0}$). On the other hand, when the
bifurcation is associated with $k=0$ the new state is spatially uniform
(homogeneous instability), while if $k\neq 0$ the instability is pattern
forming. In both the linear and quadratic models the expressions for the $%
c_{n}\left( k^{2}\right) $ coefficients are simple enough as to allow us to
locate all the static instabilities analytically, while the Hopf
instabilities can be efficiently and systematically found numerically, and
even analytically in the experimentally relevant limits $\gamma \ll 1$ and $%
\Omega \gg 1$. We will give further details in future works.

\textbf{Details of the numerical simulation of the model equations.} As explained in the text, we have performed an extensive numerical analysis of the patterns appearing in the model equations (\ref{model_norm}) of the main Letter, and here we want to say a few words about the method we have used, as well as show some examples of the transverse structures which we have found.

We have performed the numerical simulation of the evolution equations by using a symmetrized split-step Fourier method, whose name comes from the fact that it treats the linear and non-linear terms of the evolu- tion equations separately, alternating them in short time steps. The linear evolution is computed in the spatial frequency domain by using the fast Fourier transform;
as for the evolution coming from the non-linear terms, it
turns out that we can solve it exactly at every step for
these model equations. The method is exact up to second
order in the (normalized) time step, and given that we
move to Fourier domain in the spatial variables, it natu-
rally requires periodic boundary conditions in the chosen
spatial window (although fixed boundary conditions can
be simulated as well, simply by forcing the fields to have
the desired values, typically zero, at the boundaries, as
we do in the last section).

We have carried simulations of the model equations both in 1D and 2D, finding periodic patterns (which turn out to be hexagons in 2D) and localized structures which satisfy all the properties of cavity solitons (e.g., they can be written and erased individially). In Fig. \ref{Fig2D} we show examples of such structures as obtained from the linear
model, Eqs. (\ref{model_norm}) with $\mu=1$ in the main text. From a nu-
merical point of view, in order to find periodic patterns,
the reciprocal spatial lattice which we choose has to con-
tain the wave vectors which form such pattern. In the
case of the localized structures, the spatial window has
to be large enough to hold the number of solitons which
we want to write, as well as allow for their mobility if
that’s the case.

\textbf{Details about the numerical simulation of the proposed
implementation. }In the main text we showed two examples of structures
appearing when numerically simulating the quasi-1D implementation explained
in the last section, what indeed proves that patterns can be observed
with it; let us here give details about the actual parameters used for such
simulations, although, as explained in the text, patterns can be easily
found wherever they are expected to appear in the bifurcation diagram.

Let us first remark that in this case we have not assumed the existence of a
prior homogeneous mode, that is, we have numerically simulated Eq. (\ref%
{model_norm}), but setting to zero the term $\Omega ^{2}Z$ in (\ref{dZ}),
leading to the equations 
\begin{subequations}
\label{EqsReal}
\begin{gather}
\partial _{\tau }F=\left[ -1+\mathrm{i}\left( \Delta +\nabla ^{2}+Z^{\mu
}\right) \right] F+E, \\
\partial _{\tau }^{2}Z+\gamma \partial _{\tau }Z-\Omega ^{2}\rho ^{2}\nabla
^{2}Z=\Omega ^{2}Z^{\mu -1}\left\vert F\right\vert ^{2},
\end{gather}
\end{subequations}
with $\mu =1$ or $2$ for the linear or quadratic models, respectively. Note
that in both sets of equations $\Omega $ can be eliminated with the change $%
\Omega \rho \rightarrow \rho $, $\Omega F\rightarrow F$, and $\Omega
E\rightarrow E$, but we have decided to keep it just so the definition of
normalized fields and parameters is the same as in the main text. This
equation is supplemented with fixed boundary conditions at the rectangular
frame of the membrane, that is, $Z(\pm \bar{L}_{x}/2,\bar{y})=0=Z(\bar{x}%
,\pm \bar{L}_{y}/2)$, which we directly impose in the split-step method. Of
course, $\bar{L}_{x}$ and $\bar{L}_{y}$ are the dimensions of the membrane
along the $x$ and $y$ directions, respectively, normalized to the
diffraction length $l_{\mathrm{c}}$.

In Fig. 2(e) of the main text we show a soliton obtained from the linear
model ($\mu =1$), expected to appear wherever bistability is present and the
upper branch is unstable because of the static pattern forming bifurcation,
see Fig. 2(a). In order to be in this region, we have chosen the following
parameters in (\ref{EqsReal}): $\Omega =10$, $\gamma =0.1$, $\rho =1$, $%
\Delta =-2.2$. In addition we have chosen widths $\bar{L}_{y}=3.125$ and $%
\bar{L}_{x}=40$ for the membrane, and a quasi-plane injection of finite
width given by the supergaussian profile $E(\mathbf{\bar{r}})=E_{0}\exp [-(%
\bar{x}^{20}/2\sigma _{x}^{20})-(\bar{y}^{20}/2\sigma _{y}^{20})]$, with $%
\sigma _{x}=7$, $\sigma _{y}=3$, and $E_{0}=1.45$. Note that for these
choices, the effective 1D parameters become $\rho _\mathrm{eff}=3.125/\pi
\approx 0.995$ and $\Omega_\mathrm{m,eff}/\gamma_\mathrm{c}=\Omega/\rho_\mathrm{eff}\approx
10.05$, according to the simple analysis when only the fundamental mode is assumed to be excited in the $y$ axis.

On the other hand, in Fig. 2(f) of the main text we show a domain wall
expected to appear in the quadratic model ($\mu =2$) when the nontrivial
homogeneous stationary solutions with opposite signs coexist, see Fig. 2(d).
In this case we have simulated equations (\ref{EqsReal}), choosing $\Omega =%
\sqrt{0.1}$, $\gamma =0.1$, $\rho =1$, $\Delta =5$, $\bar{L}_{y}=6.25$ and $%
\bar{L}_{x}=80$, and the same kind of supergaussian illumination as before,
but with $\sigma _{x}=7.5$, $\sigma _{y}=3$, and $E_{0}=6.8$. These
parameters lead to effective 1D ones $\rho_\mathrm{eff}=6.25/\pi \approx 2$
and $\Omega_\mathrm{m,eff}/\gamma_\mathrm{c}=\Omega/\rho_\mathrm{eff}\approx 0.16$. Let us
remark that we haven't chosen a larger value of $\Omega $ because otherwise
the Hopf bifurcation tends to the point where the trivial and nontrivial
solutions connect, making it impossible to find stationary solutions above
that point (although it is possible to find dynamic ones, such as pulsing
domain walls).

\end{document}